\documentclass[aps,pra,showpacs,superscriptaddress,twocolumn]{revtex4-1}

\usepackage{amsmath,amssymb,amsthm}
\usepackage{graphicx}% Include figure files
\begin{document}

\title{Revival times at quantum phase transitions}

\author{F. de los Santos}
\affiliation{Instituto Carlos I de F{\'\i}sica Te\'orica y
Computacional, Universidad de Granada, Fuentenueva s/n, 18071 Granada,
Spain}
\affiliation{Departamento de Electromagnetismo y F{\'\i}sica de la
Materia and Instituto Carlos I de F{\'\i}sica Te\'orica y Computacional
Universidad de Granada, Fuentenueva s/n, 18071 Granada,
Spain}

\author{E. Romera}
\affiliation{Instituto Carlos I de F{\'\i}sica Te\'orica y
Computacional, Universidad de Granada, Fuentenueva s/n, 18071 Granada,
Spain}
\affiliation{Departamento de F\'isica At\'omica, Molecular y Nuclear and 
Instituto Carlos I de F{\'\i}sica Te\'orica y Computacional,
Universidad de Granada, Fuentenueva s/n, 18071 Granada, Spain
}

\date{\today}

\begin{abstract}
The concept of quantum revivals is extended to many-body systems and the
implications of traversing a quantum phase transition are explored.
By analyzing two different models, the vibron model for the bending 
of polyatomic molecules and the Dicke model for a quantum radiation field
interacting with a system of two-level atoms,
we show evidence of revival behavior for wave packets centered around energy levels
as low as the fundamental state. Away from criticality, revival times exhibit smooth, nonsingular behavior, 
and are proportional to the system size. Upon approaching a quantum critical point, they
diverge as a power law and scale with the system size, although the scaling is no longer linear.  
\end{abstract}

\maketitle
\section{Introduction}

The concept of a phase transition can be extended to zero absolute 
temperatures, when thermal fluctuations cease. Quantum fluctuations
then take over and the system may undergo a quantum phase
transition (QPT), which reflects in a dramatic change in its
physical properties as illustrated by the dependence of many several 
observables on a suitable control parameter (other than the temperature, say $\lambda$) that
determines the amplitude of quantum fluctuations \cite{qpt}. It is a generic feature 
that at second order QPTs the system will exhibit
diverging quantities as the transition is approached.
An important one is the correlation length, which, following notions and a nomenclature borrowed 
from classical critical phenomena, diverges as $\xi \sim  |\lambda-\lambda_c|^{-\nu}$.
Similarly, other quantities such as specific heats or susceptibilities diverge
in the same manner, defining new critical exponents which are related through
scaling relations. This is all well known and to date examples abound of 
quantum critical features of classical observables brought about by quantum fluctuations alone.
Much less studied, however, is the influence of QPTs on purely quantal properties,
lacking a classical counterpart. Here, we address the problem of {\it wave-packet revivals}
in systems exhibiting QPTs.

The long-time evolution of propagating quantum wave-packets may lead to unexpected periodic behavior. 
Initially, the wave packets evolve quasiclassically and oscillate with a classical period $T_{\rm Cl}$, 
but eventually spread out and collapse. At later times, multiples of the {\it revival time} $T_{\rm R}$, wave packets regain
their initial form and behave quasiclassically again. 
The classical period and the revival time of wave packet evolution are embodied in the first
coefficients of the Taylor expansion of the energy spectrum $E_k$ around the energy $E_{k_0}$ corresponding to the peak of the
initial wave packet,
\begin{equation}
E_k = E_{k_0} + E'_{k_0} (k- k_0)+ \frac{E''_{k_0}}{2} (k-k_0)^2 + \frac{E'''_{k_0}}{6} (k-k_0)^3  +\cdots,
\end{equation}
the first-, second- and third-order terms in the expansion
providing the classical period of motion $T_{\rm Cl} = 2\pi/|E'_{k_0}|$, the
quantum revival scale time $T_{\rm R}= 4\pi/|E''_{k_0}|$, and the so-called
super-revival time $T_{\rm SR}=12\pi/|E'''_{k_0}|$, respectively. 
Revivals have received considerable attention over the last
decades \cite{rob}. Both experimental and theoretical progress was made in, among others, Rydberg
atoms, molecular vibrational states, electric currents in graphene or Bose--Einstein condensates
\cite{BE,BES,BE1,BE2,BE3,BE4,BE5}.
Recently, methods for
isotope separation \cite{isotope}, number factorization \cite{factorization} as well as for wave-packet control
\cite{wp_control1,wp_control2,wp_control3}
have been put forward that are based on revival phenomena, and the presence of effective multi-body interactions in a system of
ultracold bosonic atoms in a three-dimensional optical lattice was identified in time-resolved traces of quantum 
phase revivals \cite{multibody}.
Interestingly, the collapse and revival dynamics of ultracold atoms
in optical lattices have been investigated and shown to be strongly sensitive to the initial many-body ground
state \cite{ultracold}.

In this article, we extend the concept of quantum revivals to many-body systems and explore the
implications of traversing a quantum phase transition.
By analyzing two different models, the U(3) vibron model for the bending dynamics of molecules and 
the Dicke model of two-level atoms interacting with a one-mode radiation field,
we show that, as a consequence of the squeezing of the energy levels around the ground-state energy at the critical
point, revival times of wave packets centered around quantum numbers
as low as the fundamental state diverge as a power law upon approaching a quantum 
critical point. Interestingly, we find that the revival times diverge 
sufficiently close to the critical point even at finite system-sizes.

\section{The U(3) vibron model}
The U(3) vibron model has been successfully applied to study the bending dynamics of 
linear polyatomic molecules \cite{iachello_oss}. Its Hamiltonian is constructed 
as a combination of invariant operators associated with the subalgebras of U(3)
(see \cite{iachello_oss} for further details), and it reads
\begin{equation}
H=(1-\chi) k +\frac{\chi}{N-1} \hat P,
\label{ham}
\end{equation}
where $0\leq \chi \leq 1$ and $\hat P$ is the so called pairing operator (see below). 
Here $N$ and $k$ are the total number of bound states and the number of vibrational quanta.
The associated base has elements $|k,l\rangle$, $l$ being the value of the angular momentum
along the $z$ axis perpendicular to the plane of vibrations. Their allowed values for a
given $N$ are $k=0, 1,2, \ldots, N$ and $l= \pm k, \pm(k-2), \ldots, \pm 1$ 
or 0 for $k$ odd or even.
In what follows we shall restrict ourselves to zero vibrational angular momentum, $l=0$.

Four cases can be distinguished in terms of the value of the control parameter $\chi$:
bending vibrations of rigidly linear molecules ($\chi=0$); the quasilinear case ($0<\chi\leq 0.2$),
(including a quantum critical point at $\chi=0.2$; see below);
the quasibent case ($0.2<\chi<1$); and the rigidly bent case ($\chi=1$). The Hamiltonian
is U(2) invariant for $\chi=0$ and SO(3) invariant for $\chi=1$, and in these cases analytical 
solutions for the spectra exist \cite{algebraic}: 
\begin{eqnarray}
E_{k}(\chi=0) &=& 2 k,\\
E_{k}(\chi=1) &=& \frac{4}{N-1}\left[\left(N+\frac{1}{2}\right)k-k^2 \right].
\end{eqnarray}
The classical and revival times can be computed from the expressions above to give
the exact, limiting values $T_{\rm Cl}(\chi=0)=\pi$ and $T_{\rm R}(0)=\infty$ on the one hand, 
and $T_{\rm Cl}(\chi=1)=(N-1)\pi/(2N+1-4k_0)$ and $T_{\rm R}(1)=(N-1) \pi/2$ on the other. 
The divergence of $T_{\rm R}$ is in accordance with the harmonic nature of the Hamiltonian 
when $\chi=0$, while $T_{\rm R}(\chi=1)$ being linear in $N$ provides an example of
the scaling of revival times in a many-body system, a scaling feature that is absent in the classical period.

For general $0<\chi<1$ the spectra are no longer analytically accessible and one has to
resort to numerical methods. To diagonalize the Hamiltonian Eq. (\ref{ham}) we use the matrix elements of 
$\hat P=N(N+1)-{\hat W}^2$ \cite{algebraic},
\begin{eqnarray}
\langle k_2, l |{\hat W}^2|k_1, l\rangle &=& a \delta_{k_1,k_2}
+b \delta_{k_1-2,k_2}+c\delta_{k_1+2,k_2}, 
%a(N,k_1) &=& (N-k_1)(k_1+2)+(N-k_1+1)k_1+l^2, \nonumber \\
%b(N,k_1) &=&-\sqrt{(N-k_1+2)(N-k_1+1)(k_1+l)(k_1-l)}, \nonumber \\
%c(N,k_1) &=&-\sqrt{(N-k_1)(N-k_1-1)(k_1+l+2)(k_1-l+2)}.
%&&\left[(N-k_1)(k_1+2)+(N-k_1+1)k_1+l^2\right] \delta_{k_1,k_2} \nonumber \\
%&&-\sqrt{(N-k_1+2)(N-k_1+1)(k_1+l)(k_1-l)} \delta_{k_1-2,k_2} \nonumber \\ 
%&&-\sqrt{(N-k_1)(N-k_1-1)(k_1+l+2)(k_1-l+2)} \delta_{k_1+2,k_2}. \nonumber \\ 
\end{eqnarray}
with 
\begin{eqnarray}
a &=& (N-k_1)(k_1+2)+(N-k_1+1)k_1+l^2, \nonumber \\
b &=&-\sqrt{(N-k_1+2)(N-k_1+1)(k_1+l)(k_1-l)}, \nonumber \\
c &=&-\sqrt{(N-k_1)(N-k_1-1)(k_1+l+2)(k_1-l+2)}. \nonumber \\
\end{eqnarray}

\begin{figure}
\includegraphics[width=.5 \textwidth]{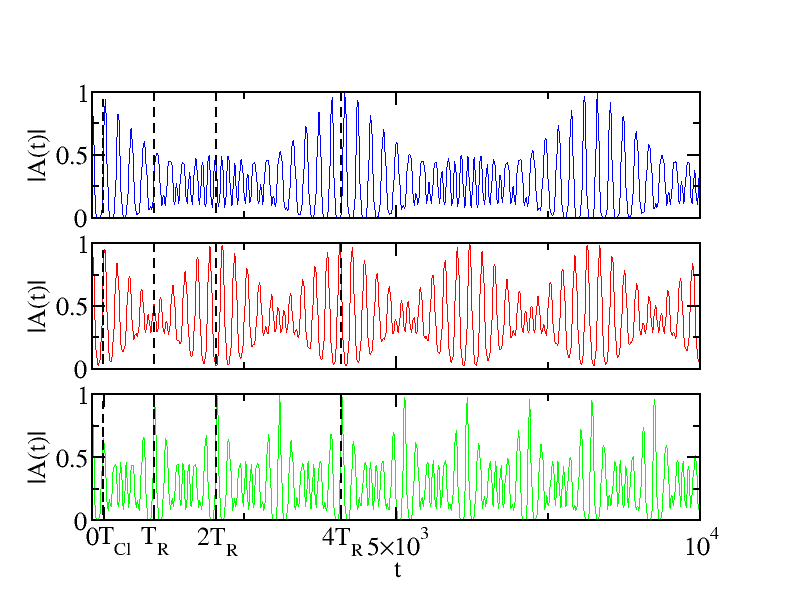}
\caption{(Color online) Time dependence of the modulus of the autocorrelation function 
$|A(t)|$ for wave packets initially centered 
around $k_0/N=0.25$ with $N=1000$ (bottom), 2000 (center), and 4000 (top).
Other parameter values are $\chi=0.5$ and $\sigma=2$. $T_R\simeq 1024$ denotes the 
revival time for $N=1000$, and $T_{\rm Cl} \simeq 192$ the classical period. Times are given
in dimensionless units.}
\label{fig1}
\end{figure}

We find the general feature (already reported in \cite{ceijnar}) 
that on increasing the system size $N$, $k/N$ becomes a quasi continuous variable $x$ and 
that the combination $\frac{E_{k/N}}{N}$ converges to a common spectrum $e_{x}$. This property 
alone implies that the revival times for a given $x$ scale with $N$ because
\begin{equation}
\frac{d^2 e_x}{dx^2}= N\frac{d^2 E_k}{dk^2}.
\label{quasicontinuous}
\end{equation}
For the purpose of illustrating this scaling, we construct initial wave packets as the linear
combination 
\begin{equation}
|\Psi(t=0) \rangle=\sum_k c_k |k,l=0\rangle
\end{equation}
with Gaussian coefficients $c_k\propto \exp[-(k-k_0)^2/\sigma]$, $\sigma=2$ and centered around $x_0=k_0/N=0.25$. 
Centering the wave-packets around a common $x_0$ 
guarantees that the average energies are common in turn \cite{note}.
At later times, of course, 
\begin{equation}
 |\Psi(t)\rangle = \sum_k c_k |k,l=0\rangle e^{-iE_k t}.
\end{equation}
Figure \ref{fig1} shows the time evolution of the modulus of the autocorrelation function, $A(t)=\langle \Psi(0) |
\Psi(t)\rangle$, which is the overlap between the initial and the time-evolving wave packet, 
of three wave-packets corresponding to $\chi=0.5$, $N=4000, 2000$, and 1000. Given an initial state, $|A͑(t)|$ decreases
in time and the occurrence of revivals is reflected in its returning to its initial value of unity.

It can be clearly appreciated from the figure that $T_{\rm R}(N=4000) \simeq 2T_{\rm R}(N=2000) \simeq 4T_{\rm R}(N=1000)$.
For consistency, we have also verified that these revival times match those obtained by evaluating $4\pi/|E''_{k_0}|$,
the second derivatives being simply computed through the numerical approximation $E_{k+1}+E_{k-1}-2E_k$.
The figure also shows the classical period $T_{\rm Cl} \simeq 192$, which does not scale with $N$. 

\begin{figure}
\includegraphics[width=.5 \textwidth]{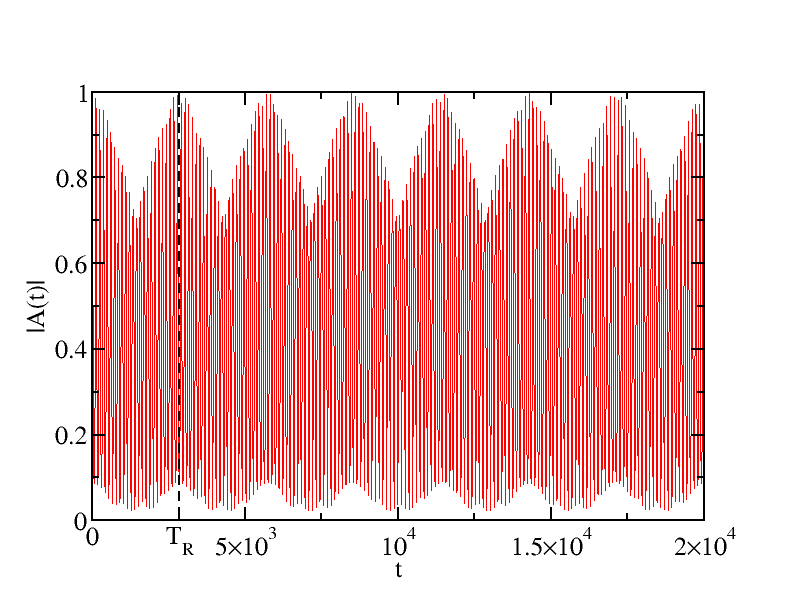}
\caption{(Color online) Time dependence of $|A(t)|$ for a wave-packet centered
around the ground state and system parameters $\chi=0.5$, $N=2000$ and $\sigma=2$. 
$T_{\rm R}\simeq 2853.5$ denotes the revival time. Time is given
in dimensionless units.}
\label{fig2}
\end{figure}

Revivals are also observed at energy levels as low as the ground state. As an example, Fig. \ref{fig2} shows the time
development of a wave packet with Gaussian-distributed population for $\chi=0.5$, centered around the ground state and with
$\sigma=2$. 
The estimated revival time from the figure is approximately 2850, to be compared with $T_{\rm R}/2=2\pi/|E''_{k_0=0}|=2853.5$.
In this case, $E''_k$ is computed by fitting the first three levels of the spectrum to a parabola and then taking $k=0$.
The results thus obtained are in perfect agreement with those observed by monitoring the wave-packets time evolution.
Other values of $\chi$, both above and below 0.2, yield again the behavior $T_{\rm R} \sim N$.

\begin{figure}
\includegraphics[width=.5 \textwidth]{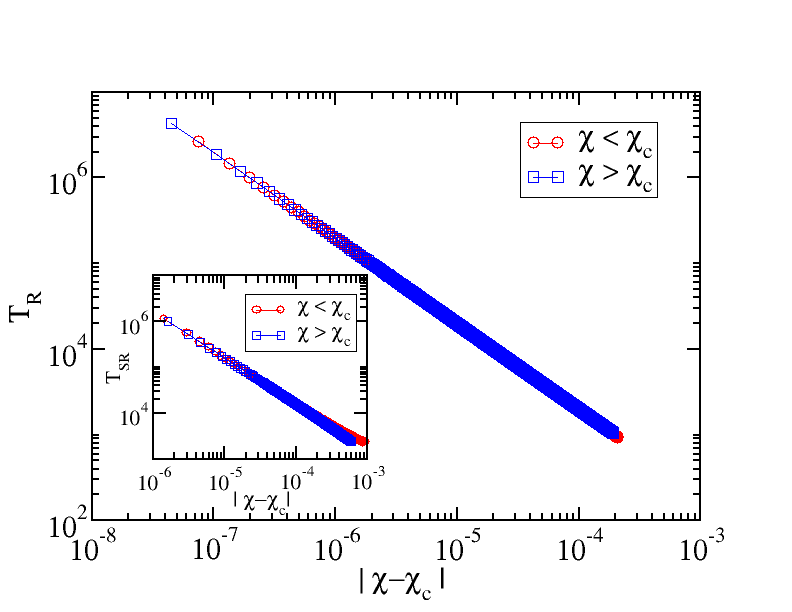}
\caption{(Color online) Revival and super-revival (inset) times at criticality as a function of $\chi$ for wave-packets centered 
around the ground state and $N=1000$. The red circles and the blue squares correspond to, respectively, 
$\chi < \chi_c$ and $\chi > \chi_c$. Times are given in dimensionless units.}
\label{tr_tsr_k0}
\end{figure}

Next, we study how the revival times are affected by the presence of the quantum critical point. 
At $\chi=\chi_c=0.2$ this system undergoes a second-order quantum phase transition in the thermodynamic
limit, $N \to \infty$, between two phases displaying anharmonicities of opposite signs.
To compute the classical period and the revival time, notice that the former is simply related to the energy gap, 
$\Delta \equiv E_1-E_0$, 
by $T_{\rm Cl} \sim \Delta^{-1}$. Since this can be evaluated in the thermodynamic limit through
$\Delta=\sqrt{(5\chi-1)(1+3\chi)}$ (valid for $\chi>\chi_c$
\cite{finite_size}), this implies that the classical period of wave-packets centered around the ground state diverges
as the critical point is approached as $T_{\rm Cl} \sim (\chi-\chi_c)^{-1/2}$. For the sake of consistency, we have verified this
prediction numerically.
As regards $T_{\rm R}$, a divergent behavior is also expected 
due to the squeezing of the energy levels around the ground state energy at the critical point,
what leads to the time evolution of the wave-packet being basically controlled by a single autostate.
In fact, we find $T_{\rm R} \sim (\chi-\chi_c)^{-1}$ at fixed $N=1000$ and $k_0=0$, with $\chi_c = 0.205 \, 907 \, 075(3)$ (see
Fig.
\ref{tr_tsr_k0}).
Supporting this image is the divergence of the super-revival time, $T_{\rm SR}$, as shown in the inset of Fig. \ref{tr_tsr_k0}
with, in this case, $\chi_c = 0.203 \,904 \,4(2)$ and the same scaling exponent --1. Notice, that all time scales diverge at close
but
different $\chi$, only to coincide at $\chi_c$ in the thermodynamic limit [for $N=1000$, $T_{\rm Cl}$ peaks at $\chi_c =
0.205 \, 305(5)$ rather than diverging]. 

Another interesting aspect is the dependence of $T_{\rm R}$ and $T_{\rm Cl}$ on the system-size $N$ for a given $k$.
Previous results for the energy gap yield the scaling form $\Delta \sim N^{-1/3}$ (see \cite{algebraic} and references
therein), which is in perfect agreement with our own, $T_{\rm Cl}(k_0=0)\sim N^{1/3}$. 
Moreover, we find this same scaling behavior for wave-packets peaked at arbitrarily high 
$k$, i.e. $T_{\rm Cl}(k_0 >0)\sim N^{1/3}$. 
\begin{figure}
\includegraphics[width=.5 \textwidth]{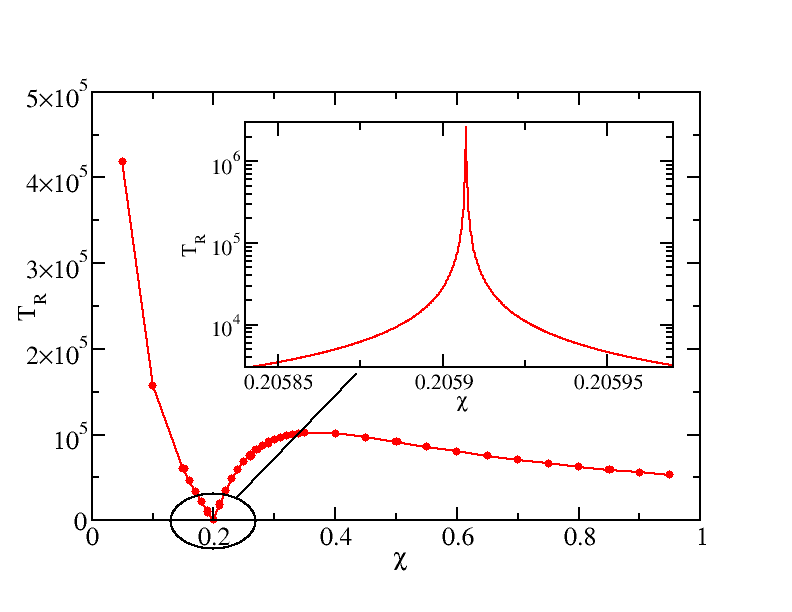}
\caption{(Color online) Revival time as a function of $\chi$ for wave-packets centered around the ground state
and $N=1000$. The inset is an enlargement of the delimited area and shows the divergence of $T_R$
close to the critical point. Notice the logarithmic scale of the $y$ axis. Time is given
in dimensionless units.}
\label{fig3}
\end{figure}
Turning to $T_{\rm R}$, we find that for any system-size $N$ there always 
exists a $\chi$, the closer to $\chi_c$ the larger $N$, such that $T_{\rm R}(k_0=0)$ diverges. 
This is illustrated in the inset of Fig. \ref{fig3} which shows the behavior of $T_R$ for $N=1000$ 
in the vicinity of $\chi_c$. Similar divergences are found generically for any $N$, including values as low as 10,
and for a different quantity in another model too, as discussed in the next section.
To rationalize this behavior one can resort to analytical approximations to the spectra. 
In particular, a semiclassical approach at $\chi_c$ gives \cite{ceijnar}
\begin{equation}
\frac{E_k}{N} -\chi_c \sim \left( \frac{k}{N}\right)^{4/3},
\end{equation}
which yields $T_{\rm R} \sim k_0^{2/3} N^{1/3}$ and $T_{\rm Cl}\sim k_0^{-1/3}N^{1/3}$,
$k_0$ being again the quantum number of the energy level the packet is centered around. 
However, this approximation seems to be valid only
far from the ground state, as shown in Fig. \ref{fig4}
by plotting $\frac{E_k}{N} -\chi_c$
vs. $k$ in double-logarithmic scale for several system sizes. Note that the spectra compare well
with the power law $k^{1.36}$ only for sufficiently high $k$. 
This conveys the idea that the semiclassical formula might be asymptotically 
correct at very large values of both $k$ and $N$.

\begin{figure}
\includegraphics[width=.5 \textwidth]{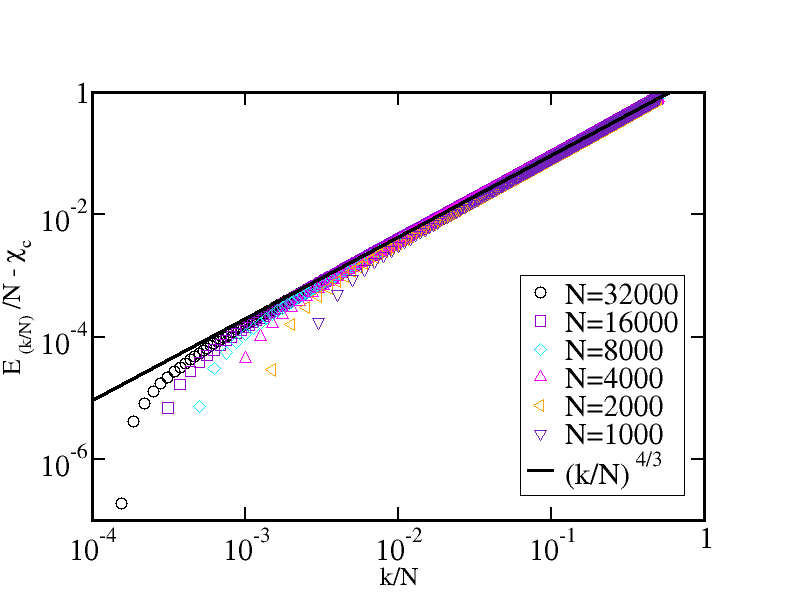}
\caption{(Color online) Numerical spectra at criticality for increasing system-sizes $N$ as compared with reults of
the semiclassical
approximation. (Dimensionless units.)}
\label{fig4}
\end{figure}

\section{The Dicke model}
The Dicke model describes the interaction of a two-level atomic ensemble with a one-mode 
radiation field. It dates back to the 1950s \cite{dicke} and to date several experimental 
realizations of the model have been proposed \cite{garraway}. 
The Dicke model exhibits a quantum phase transition at zero temperature
as embodied in the Hamiltonian \cite{emarybrandes}
\begin{equation}
 H=w_0 J_z +w a^\dagger a + \frac{\lambda}{\sqrt{2j}} (a^\dagger +a)(J_++J_-).
\end{equation}
Here, $J_z$ and $J$ are the usual angular momentum operators
for collective spin operators of length $j=N/2$, and $a$, $a^\dagger$ are the
bosonic operators of the field. The atomic level splitting is
given by $w_0$, $w$ is the field frequency, and $\lambda$ is the atom-field coupling. 
In the thermodynamic limit, $N, j \to \infty$, the system undergoes a second-order quantum phase
transition at a critical coupling of $\lambda=\lambda_c=\sqrt{w_0w}/2$. At this
point the system changes from a normal phase to the so-called
super-radiant one in which both the field and
the atomic ensemble acquire macroscopic occupations.
Revivals and fractional revivals in the Jaynes-Cummings model, a {\em single} ($j=1/2$) two-level atom
interacting with a one mode of the quantized radiation field, were studied in the past 
\cite{jaynes-cummings,jaynes-cummings2}.

\begin{figure}
\includegraphics[width=.5 \textwidth]{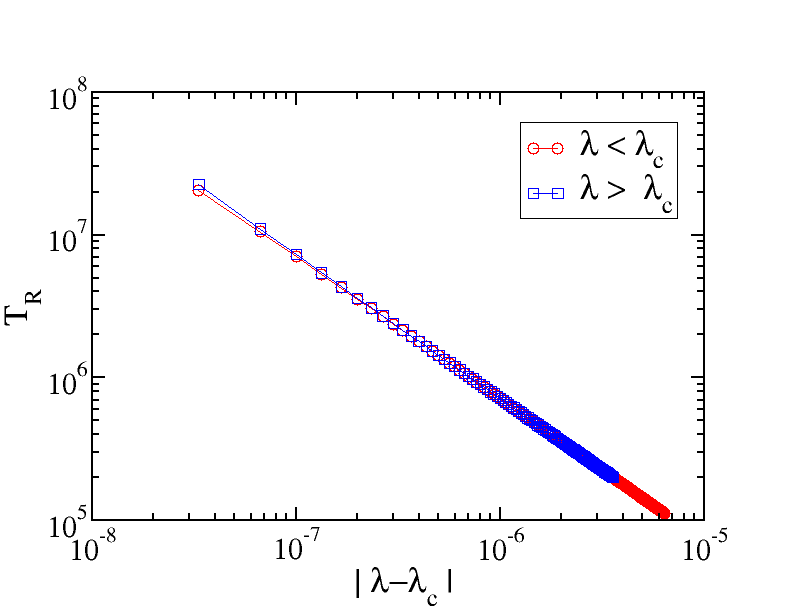}
\caption{(Color online) Log-log plot of the revival times
for the Dicke model with $j=10$ in the vicinity of 
the critical point. The red circles and the blue squares correspond to, respectively, 
$\lambda < \lambda_c$ and $\lambda > \lambda_c$. The data can approximated by a straight line
of slope --1. (Atomic units.)} 
\label{dicke}
\end{figure}

Proceeding along the same lines as in the previous section, that is, 
solving numerically for the spectra and populating the autostates around 
the ground state with Gaussian weights, a wave-packet is constructed and
its time development monitored. The parity has been taken into account as a
symmetry in this system. (Variational approximations have been proposed in 
\cite{casta1,casta2} to study the
Dicke model in an analytical framework). We have verified that our results for 
the classical time on scale resonance, i.e. $w_0=w=1$, follow those 
reported in \cite{vidal}, $\Delta \sim (\lambda_c-\lambda)^{1/2}$.
As for $T_{\rm R}$, we find $T_{\rm R}\sim | \lambda_c-\lambda|^{-1}$ (see Fig. \ref{dicke}).
Oddly enough, as in the vibron model, there is a value of $\lambda$
at which $T_{\rm R}$ diverges irrespective of the system-size.
At this point, it may be in order to mention that it has been recently shown that the fidelity 
necessarily diverges at the critical point in the Dicke model \cite{casati}, and, interestingly, 
divergent behavior has been observed to occur at finite system-sizes \cite{hirsh}.

\section{Conclusions}
We have extended the concept of quantum revivals to many-body systems 
and explored the implications of traversing a quantum phase transition by analyzing
the time development of wave-packets centered around the fundamental state in two different models, 
namely, the U(3) vibron model and the Dicke model. Far from the quantum critical point,
characteristic time scales such as the classical period and the revival and the superrevival times
exhibit a smooth non-singular behavior, and revival times are proportional to 
the system size. Upon approaching a quantum critical point, energy levels squeeze 
into the ground state, rendering the wave packet with a Gaussian-distributed population basically 
a combination of states with almost  
equal energies. We have shown evidence that under these circumstances, the above-mentioned quantities 
diverge as power laws with well-defined critical exponents. Interestingly, the revival and super-revival 
times were found to diverge on approaching the critical point even at finite system size.
Should it be the case that {\em all} time scales in the Taylor expansion diverge, a statement we cannot confirm but 
that is suggested by the numerics and by the squeezing of the energy levels around the ground state as the transition point is 
approached, the time evolution of such wave packets at criticality would then be limited to phase changes, i.e., to rotations. 
Finally, we comment that quantum phase transitions can also influence the revival behavior of wave packets 
centered around excited states. This interesting effect, however, will be discussed elsewhere.

\begin{acknowledgments}
This work  was supported by the Spanish Projects No. MICINN FIS2009-08451,
No. FQM-02725 (Junta de Andaluc\'ia), No. 20F12.41 (CEI BioTic UGR), and No. MICINN FIS2011-24149. 
\end{acknowledgments}


\begin{thebibliography}{99}

\bibitem{qpt}
S. Sachdev,  {\it Quantum Phase Transitions}
(Cambridge University Press, New York, 1999).

\bibitem{rob} 
R.W. Robinett,
%Quantum wave packet revivals 
Phys. Rep. {\bf 392}, 1 (2004).

\bibitem{BE} 
E. Romera and F. de los Santos 
%Revivals, classical periodicity, and zitterbewegung of electron currents in monolayer graphene
Phys. Rev. B {\bf 80}, 165416 (2009).

\bibitem{BES}
A. L\'opez, Z.Z. Sun, and J. Schliemann,
Phys. Rev. B {\bf 85}, 205428 (2012)

\bibitem{BE1}
G. Rempe, H. Walther, and N. Klein,
Phys. Rev. Lett. {\bf 58}, 353 (1987).

\bibitem{BE2} 
J.A. Yeazell, M. Mallalieu, and C.R. Stroud Jr.,
Phys. Rev. Lett. {\bf 64}, 2007 (1990).

\bibitem{BE3} 
T. Baumert, V. Engel, C.  R\"ottgermann, W.T. Strunz, and G. Gerber,
%Femtosecond pump—probe study of the spreading and recurrence of a vibrational wave packet in Na$_2$
Chem. Phys. Lett. {\bf 191}, 639 (1992).

\bibitem{BE4} 
M.J.J. Vrakking, D.M. Villeneuve, and A. Stolow,
Phys. Rev. A {\bf 54}, R37-R40 (1996).

\bibitem{BE5} 
A. Rudenko, Th. Ergler, B. Feuerstein, K. Zrost, C.D. Schr\"oter, R. Moshammer, and J. Ullrich,
%Real-time observation of vibrational revival in the fastest molecular system 
Chem. Phys. {\bf 329}, 193  (2006).


\bibitem{isotope}
I.-Sh. Averbukh, M.J.J. Vrakking, D.M. Villeneuve, and A. Stolow,
Phys. Rev. Lett. {\bf 77}, 3518 (1996).

\bibitem{factorization}
M. Mehring, K. Mueller, I.-Sh. Averbukh, W. Merkel, and W.P. Schleich,
Phys. Rev. Lett. {\bf 98}, 120502 (2007). 

\bibitem{wp_control1}
E.A. Shapiro, M. Spanner, and M.Y. Ivanov,
Phys. Rev. Lett. {\bf 91}, 237901 (2003).

\bibitem{wp_control2} 
M. Spanner, E.A. Shapiro, and M.Y. Ivanov, 
Phys. Rev. Lett. {\bf 92}, 093001 (2004).

\bibitem{wp_control3}
K.F. Lee, D.M. Villeneuve, P.B. Corkum, and E.A. Shapiro,
Phys. Rev. Lett. {\bf 93}, 233601 (2004).

\bibitem{multibody}
S. Will, T. Best, U. Schneider, L. Hackermuller, D.S. Luhmann, and I. Bloch,
Nature (London) {\bf 465}, 197 (2010).

\bibitem{ultracold} 
E. Tiesinga and P.R. Johnson,
Phys. Rev. A {\bf 83}, 063609 (2011).

\bibitem{iachello_oss}
F. Iachello and S. Oss,
J. Chem. Phys. {\bf 104}, 6956 (1996).

\bibitem{algebraic}
F. P\'erez-Bernal and F. Iachello,
Phys. Rev. A {\bf 77}, 032115 (2008).

\bibitem{ceijnar}
M.A. Caprio, P. Cejnar, and F. Iachello,
Ann. Phys. (N.Y.) {\bf 323}, 1106-1135 (2008).

\bibitem{note}
In principle, $\sigma$ should be chosen such that the energy variances for different $N$ are comparable, but we have checked that
this is not a relevant issue.

\bibitem{finite_size}
P. P\'erez-Fern\'andez, J.M. Arias, J.E. Garc\'ia-Ramos, and F. P\'erez-Bernal,
Phys. Rev. A {\bf 83}, 062125 (2011).

\bibitem{dicke}
R.H. Dicke, 
Phys. Rev. {\bf 93}, 99 (1954͒).

\bibitem{garraway}
B.M. Garraway, 
Philos. Trans. R. Soc. A {\bf 369}, 1137-1155 (2011).

\bibitem{emarybrandes}
C. Emary and T. Brandes, 
Phys. Rev. E {\bf 67}, 066203 (2003).

\bibitem{jaynes-cummings}
J. Gea-Banacloche,
Phys. Rev. Lett. {\bf 65}, 3385 (1990).

\bibitem{jaynes-cummings2}
I.-Sh. Averbukh,
Phys. Rev. A {\bf 46}, R2205 (1992).

\bibitem{casta1}  
O. Casta\~nos,  E. Nahmad-Achar, R. L\'opez-Pe\~na, and J.G. Hirsch,
Phys. Rev. A {\bf 83}, 051601 (2011).

\bibitem{casta2}  
O. Casta\~nos, E. Nahmad-Achar, R. L\'opez-Pe\~na, and J.G. Hirsch,
Phys. Rev. A {\bf 84}, 013819 (2011).

\bibitem{vidal}
J. Vidal and S. Dusuel, 
Europhys. Lett. {\bf 74}, 817 (2006).

\bibitem{casati}
O. Casta\~nos, E. Nahmad-Achar, R. L\'opez-Pe\~na, and J.G. Hirsch,
Phys. Rev. A {\bf 86}, 023814 (2012). 


\bibitem{hirsh}
W.G. Wang, P. Qin, Q. Wang, G. Benenti, and G. Casati 
Phys. Rev. E {\bf 86}, 021124 (2012).


\end{thebibliography}
\end{document}